\documentclass[sigconf,natbib=false, nonacm]{acmart}
\AtBeginDocument{%
  }

\RequirePackage[
  datamodel=acmdatamodel,
  style=acmnumeric,
  ]{biblatex}

\begin{document}

\title{Utilizing LLMs for Industrial Process Automation}


\author{Salim Fares}
\affiliation{%
\institution{University of Passau}
\city{Passau}
\country{Germany}}
\email{salim.fares@uni-passau.de}

\renewcommand{\shortauthors}{Fares et al.}

\begin{abstract}
\label{sec:abstract}
A growing number of publications address the best practices to use Large Language Models (LLMs) for software engineering in recent years. However, most of this work focuses on widely-used general purpose programming languages like Python due to their widespread usage training data. The utility of LLMs for software within the industrial process automation domain, with highly-specialized languages that are typically only used in proprietary contexts, remains underexplored. This research aims to utilize and integrate LLMs in the industrial development process, solving real-life programming tasks (e.g., generating a movement routine for a robotic arm) and accelerating the development cycles of manufacturing systems.
\end{abstract}



\received{20 February 2007}
\received[revised]{12 March 2009}
\received[accepted]{5 June 2009}

\maketitle


\section{Introduction}
\label{sec:introduction}
In Industrial Process Automation (IPA), specialized software is developed in proprietary programming languages ---e.g. Programmable Logic Controller (PLC), ABB’s RAPID Programming Language (RAPID)--- to control complex manufacturing machines. This programming is vendor-dependent~\cite{yang2025autoplcgeneratingvendorawarestructured} and is not supported by commercial Artificial Intelligence (AI) systems for software development~\cite{Lee2025} such as GitHub Co-Pilot~\cite{pandey2024transformingsoftwaredevelopmentevaluating}. 
LLMs like GPT~\cite{openai2024gpt4technicalreport} and Llama~\cite{grattafiori2024llama3herdmodels} are trained on massive, open, and diverse text datasets, code, documents, web pages, etc. In comparison, the IPA utilizes data such as schedules that define the timing and sequence of operations within the automated process~\cite{ISA88}, electronic plans that define the physical interconnection of the industrial system components like sensors, power supplies and robotic arms~\cite{SiemensTIA} and functional diagrams as graphical representations of the control logic~\cite{SiemensFBD} (i.e., how the inputs lead to the outputs).

This data is proprietary, closed and cannot be shared openly by companies as their customers generally keep their data private~\cite{zheng2020privacy}. Furthermore, it is heterogeneous with many formats like XML and technical drawings~\cite{inbook}, and it is mainly formal rather than natural, i.e., symbols and wiring have technical relationships that normal LLM tokenization doesn’t capture~\cite{doris2024designqamultimodalbenchmarkevaluating}. This makes training and fine-tuning LLMs on industrial data very difficult and laborious~\cite{Klaeger_2021}. Furthermore, Small and Medium-sized Enterprises (SMEs) have small project-specific data in inconsistent formats and lack the staff to curate or annotate training datasets. They also lack both computational resources and AI expertise to develop such LLMs. Large companies
like Siemens AG which have already developed an industrial AI-assistant ~\cite{siemens_industrial_copilot}. However, it is also a closed, proprietary system with no access to the underlying model, restricting experimentation and fine-tuning. Furthermore, its tight integration with Siemens’ software creates vendor dependency and makes compatibility with non-Siemens software difficult. Even Siemens continues to face challenges in accessing data from external providers. Moreover, when data is available, differing formats, interface standards, and ecosystem constraints often require additional customization, middleware, or manual intervention to ensure smooth data exchange.

This research will help SMEs, to use their in-house data to develop AI for IPA and accelerate the development cycles of their manufacturing systems.
The Main Research Question (MRQ) addressed in this work is: 

\begin{description}

\item[MRQ:] \textit{How can Large Language Models (LLMs) be adapted and integrated to generate and optimize proprietary programming languages?}
By addressing this question, we should be able to narrow the gap between mainstream LLMs and the IPA.
\end{description}
We structure our research regarding the MRQ by studying the following Research Questions (RQs):

\begin{description}
\item[RQ1:] 
\textit{What are the limitations of existing LLMs in generating industrial proprietary programming code?}
This question will help us explore the challenges, that current LLMs face dealing with proprietary programming tasks. Investigating these challenges will guide us to find whether a suitable solution is currently possible.
\item[RQ2:] 
\textit{Can a general-purpose LLM generate valid industrial proprietary code using only prompt engineering?}
With this question, we explore whether an SME on its own can already utilize existing general-purpose LLMs to assist their IPA without expensive solutions like training their own models. To this end, we explore the hypothesis that current LLMs can be used with carefully engineering few-shot prompts to support specific, simple but often reoccurring, programming tasks.

\item[RQ3:] 
\textit{How to integrate and utilize the different data modalities to assist and/or train an LLM to generate accurate and functional IPA code?}
We explore different data modalities that are used in IPA like schedules, electronic plans, functional diagrams. Out of those, we create a local data pipeline, choosing the required data blocks and integrating the different modalities for LLMs to generate the correct and desired program.

\end{description} 

\section{Related Work}
\label{sec:related_work}
Recent studies have increasingly examined how AI and LLMs can support industrial programming. Fakih~\textit{et al.}~\cite{Fakih_2024} introduced a user-guided iterative framework, Large Language Model for Programmable Logic Controller (LLM4PLC), that utilizes LLMs to translate natural language requirements into PLC code. To further enhance performance, they applied Low-Rank Adaptation (LoRA) fine-tuning and prompt engineering techniques. Evaluation results demonstrated a notable increase in the $pass@k$ metric (i.e., the rate of successful code generation, as defined in \cite{chen2021evaluatinglargelanguagemodels}. Additionally, a group of PLC programming experts assessed the generated code on a 1–10 scale according to predefined factors: correctness, maintainability, and compliance with industry best practices.

Similarly, Antero~\textit{et al.}~\cite{robotics13090137} developed an approach aimed at lowering the cost of programming complex robotic behaviors using LLMs. Their framework employed a Generator LLM to create task plans composed of predefined, human-written software blocks structured as Finite State Machines (FSMs). These plans were then examined by a Supervisory LLM, which verified their accuracy and proposed corrections when necessary. This generator–supervisor cycle continued until an error-free plan was achieved or the iteration limit was reached, after which the Generator produced a JSON-formatted representation of the final plan. The Generator was also provided with contextual details about the environment, nearby objects, their attributes, and permissible actions, allowing it to determine how to reach a user-specified goal state using the available resources. The method was tested on 11 manually designed multi-action tasks, successfully producing valid plans for 8 of them.

Morano-Okuno~\textit{et al.}~\cite{10823586} propose a framework that blends digital twins, AI-driven reasoning, and human–robot interaction models to make collaborative robot (cobot) programming more accessible to non-experts. In their system, users train cobots interactively inside a virtual environment. The AI interprets the simulated workspace, recommends and adjusts task plans built from predefined actions, integrates user input, and repeatedly tests and refines behaviors—without producing executable code. The authors demonstrate the framework through several scenario-based simulations, but they do not provide detailed quantitative, task-level performance measures (e.g., success rates or time-to-completion). Instead, their evaluation focuses on how interactive simulation lowers the learning barrier and improves usability, rather than on generating or assessing deployable code.

In contrast to prior work, we do not just train an LLM to generate industrial code, but we analyze in depth the limitations of existing LLMs and explore their suitable solutions, we also investigate how to integrate the different data modalities used in the IPA.

\section{Proposed Approach}
\label{sec:proposed_approach}
We hypothesize that: 1) LLMs exhibit generalization to domain-specific languages like PLC and RAPID, even if it was not explicitly trained on them; 2) integrating the different data modalities can help LLMs to generate more accurate and efficient industrial solutions. 
We begin with prompt engineering to address RQ2, as it offers a straightforward and easily-applied strategy for resource-constrained settings like IPA. We then revisit RQ1, broadening our approach beyond prompt engineering to incorporate Retrieval-Augmented Generation (RAG) and lightweight fine-tuning methods such as LoRA. Finally, we advance to multimodal data integration to address RQ3.

\subsection{Prompt Engineering}
To answer RQ2 we test the current downstream LLMs on common, real-life tasks and explore the limitations by demonstrating the efficiency and productivity of the generated solutions and to which extent these LLMs are able to follow the proprietary standards of an SME for the provided tasks. We try to assist the LLM through prompt engineering without fine-tuning, guiding the model with the required instructions and few examples in the prompt to execute the tasks in the IPA. We test the capabilities of LLMs across tasks of diverse complexity, and record the results to analyze the obstacles these LLMs face by validating the generated program against the specified rules in the prompts. This approach provides an easy-to-apply and practical starting point for SMEs using local modules, and the discovered limitations in this step will also be part of the answer for RQ1 along with other aspects specified in Sec. \ref{sec:investigate_current_limitations}.

\subsection{Investigate Current Limitations}
\label{sec:investigate_current_limitations}
This study systematically investigates the limitations of current LLMs when applied to industrial proprietary programming tasks by utilizing in-house data and real-life tasks for assessing functional correctness and productivity, and a comparison of different solution approaches, such as prompt engineering, RAG, and lightweight fine-tuning (e.g., LoRA). Up to this point, we have focused solely on prompt engineering. However, a broader range of solutions is needed because prompt engineering alone provides only partial answers and does not fully address challenges related to data scarcity and domain specificity. Therefore, alternative approaches, such as RAG and fine-tuning techniques designed for limited data scenarios (e.g., LoRA), will be explored in future work to more comprehensively answer RQ1.

\subsection{Utilize Different Data Modalities}
Different data modalities like schedules, electronic plans, and source code are used in the IPA. For this, we define how each data modality will be processed and prepared, identify the required data blocks, determine how they should be grouped, and establish standardized formats for data import and export. Finally, we specify how these different data types will be integrated and used in combination to support project development. We investigate our hypothesis by assisting the LLMs through prompt engineering as in the previous step. We continue to test the LLMs on similar tasks as previously and analyze our findings to answer RQ3.

\section{Expected Contributions}
\label{sec:contributions}
This research aims to make the following contributions:
\begin{enumerate}
    \item An analysis on the strengths and limitations of current LLMs in proprietary programming.

    \item A proposed solution to mitigate the discovered limitations.

    \item A proposed solution to utilize different data modalities in the development of an industrial LLM.
\end{enumerate}

\section{Evaluation Plan}
\label{sec:evaluation}
The evaluation will follow a mixed-methods approach combining quantitative metrics (accuracy) using a custom validator, measuring adherence to proprietary standards. We also check for functional correctness using digital twins to execute generated code in a controlled virtual environment. Moreover, we gather feedback from professional engineers on productivity impact of using the developed LLM (compare development time and error rates with and without LLM assistance). Their feedback also includes a qualitative assessment regarding their experiences and interactions with developed system (e.g., how easy or hard it is to use the developed system).

\section{Initial Results}
\label{sec:initial_results}
We conducted a case study\cite{fares2025} that explored whether a general-purpose LLM can perform code modification tasks in RAPID using only prompt engineering, without additional training. This study partially answers RQ2, because it is only utilizing one LLM and considering specific movement routines in RAPID code, and partially answers RQ1 because it only considers prompt engineering. The structural nature of the programming language, that is similar to what a general-purpose LLM like Llama 3.1 70B~\cite{grattafiori2024llama3herdmodels} has already seen, gave us the idea for this case study. We evaluated three programming tasks: 1) Modifying arguments in movement routines; 2) adding offset instruction to a movement routine; 3) reversing movement routines. The accuracy the model achieved for each task is shown in table~\ref{tab1}. The results indicate that while LLMs can handle basic RAPID code modifications effectively, more complex transformations require domain-specific adaptation, motivating the next phase of this research to build a RAG system to assists the LLM with similar examples to the provided tasks from previously developed industrial projects.
\begin{table}[!h]
\centering
\caption{The Accuracy for each task}
\label{tab1}
\begin{tabular}{|c|c|c|}
\hline

& \multicolumn{2}{c|}{Accuracy}
\\
\hline

Task& German prompts & English prompts
\\
\hline

Arguments Modification&$99.71\%$&$99.36\%$
\\
\hline

Adding an Offset&$91.86\%$&$91.97\%$
\\
\hline

Reversing&$77.27\%$&$83.72\%$
\\
\hline

\end{tabular}
\end{table}

\section{Timeline}
The PhD project started in December 2024 under the supervision of Prof. Dr. Steffen Herbold at the University of Passau. In the second year of this project, we focus mainly on building the RAG system and fine-tuning the LLM to finish the answers for RQ1 and RQ2. In the third year, we will focus on multi-modal data integration to answer RQ3 and use the findings of our research to answer the main research question to complete the Ph.D. program in the third year.

\begin{acks}
This work was carried out as part of the PLC-GPT project, a collaboration between the University of Passau and AKE Technologies funded by the Bavarian State Ministry of Science and Arts.
\end{acks}

\printbibliography
\end{document}